\documentstyle [12pt]{article}
\title{Phase diagram of the asymmetric Hubbard model}
\author{Pavol Farka\v sovsk\'y\\
Institute  of  Experimental  Physics,  Slovak   Academy   of
Sciences\\
Watsonova 47, 040 01 Ko\v {s}ice, Slovakia}
\date{}
\begin{document}
\baselineskip=20pt
\maketitle

\begin{abstract}
The ground-state phase diagram of the asymmetric Hubbard model
is studied in one and two dimensions by a well-controlled numerical method.
The method allows to calculate directly the probabilities of particular 
phases in the approximate ground-state and thus to specify the stability 
domains corresponding to phases with the highest probabilities.
Depending on the electron filling $n$ and the magnitude of the asymmetry
$t_f/t_d$ between the hopping integrals of $f$ and $d$ electrons two
different scenarios in formation of ground states are observed. At low
electron fillings ($n\leq 1/3$) the ground states are always phase
segregated in the limit of strong asymmetry ($t_d\gg t_f$). With decreasing
asymmetry the system undergoes a transition to the phase separated state and
then to the homogeneous state. For electron fillings $n>1/3 $ and weak 
Coulomb interactions the ground state
is homogeneous for all values of asymmetry, while for intermediate and
strong interactions the system exhibits the same sequence of phase
transitions as for $n$ small. Moreover, it is shown that the segregated
phase is significantly stabilized with increasing electron filling, while
the separated phases disappear gradually from the ground-state phase
diagrams.  

\end{abstract}

\newpage

\section{Introduction}
The asymmetric Hubbard model is one of the simplest models 
for a description of correlated fermions on the lattice. 
It has been used in the literature to study a great variety of many-body 
effects in rare-earth and transition-metal compounds, of which quantum phase 
transitions, mixed-valence phenomena, charge-density waves, and electronic 
ferroelectricity are the most common examples~\cite{Lyzwa1,Batista,Yin,
Souza,Wang}. In the last few years the asymmetric Hubbard model was also 
used for a description 
of ground-state properties of fermionic particles on the optical 
lattice~\cite{Valencia,Gu}. The model consists of two species of electrons: 
heavy $f$ electrons and light $d$ electrons. The Hamiltonian of the model is
 
\begin{equation}
H=-t_d\sum_{<ij>}d^+_id_j-t_f\sum_{<ij>}f^+_if_j+U\sum_if^+_if_id^+_id_i,
\end{equation}
where $f^+_i$ $(f_i)$ and $d^+_i$ $(d_i)$
is the creation (annihilation) operator  of heavy and light electron 
at lattice site $i$.

The first two terms of (1) are the kinetic energies corresponding to
quantum-mechanical hopping of $d$ and $f$ electrons 
between the nearest neighbor sites $i$ and $j$
with hopping probabilities $t_d$ and $t_f$, respectively.
The third term represents the on-site
Coulomb interaction between the $d$ electrons with density
$n_d=\frac{1}{L}\sum_id^+_id_i$ and the $f$ electrons with density 
$n_f=\frac{1}{L}\sum_if^+_if_i$, where $L$ is the number of lattice sites. 
The model is called "asymmetric" because the hopping integrals
for $d$ and $f$ electrons may be different. Usually, the hopping 
integral of the $d$ electrons is taken to be the unit of energy
$(t_d=1)$ and the $f$-electron hopping integral is considered in the 
limit $t_f \ll 1$. This is a reason why the $d$ electrons are called
light and the $f$ electrons heavy. The Hamiltonian (1) reduces to the 
spinless Falicov-Kimball model for $t_f=0$ and to the usual one-band 
Hubbard model for $t_f=1$. Thus the asymmetric Hubbard model can we
viewed as a generalized Falicov-Kimball model and  also as a generalized 
one-band Hubbard model.

The first systematic study of ground-state properties
of the asymmetric Hubbard model has been performed by Lyzwa et al. 
using various analytical and numerical
techniques~\cite{Lyzwa1,Lyzwa1a,Fath,Lyzwa2}.
In the first paper from this series the authors studied the ground-state
properties of the one dimensional asymmetric Hubbard model by own
approximate method that allowed to treat larger clusters than accessible by
exact diagonalization technique. Their method was based on the sequence of
two steps. First, they found the lowest-energy state for every permissible
$f$-electron configuration, similarly as in the pure spinless
Falicov-Kimball model ($t_f=0$).
To do this the matrices of rank $\frac{L!}{(L-N_d)!N_d!}$ had to be
diagonalized. Second, they took the states thus found as a basis of a new
matrix of rank  $\frac{L!}{(L-N_f)!N_f!}$ that was subsequently
diagonalized. The lowest-energy eigenstate of this matrix was then used to
construct the approximate ground state. The main result obtained by the
application of this method to the asymmetric Hubbard model was that the
motion of the heavy electrons is strongly influenced by the light ones,
while the light electrons are almost unaffected by the presence of the heavy
ones. The subsequent study~\cite{Lyzwa1a} of the asymmetric Hubbard model on the
one-dimensional ring with two $f$ and two $d$ electrons showed that an
effective attraction between two $f$-electrons can be produced by
correlation effects for a certain set of the model parameters. The same
result, an effective attraction between two heavy electrons mediated by two
light electrons leading to the phase segregation, was confirmed also in two 
dimensions~\cite{Lyzwa2}. These studies showed that the phase segregation is own 
not only to the Falicov-Kimball model~\cite{Freericks}, but persists also
at finite $t_f$.  For strong interactions this result was proven rigorously
by Ueltschi~\cite{Ueltschi}. The boundary of the phase segregation/separation 
region in the $U-t_f$ plane has been calculated very recently by two 
different methods. To identify the transition boundary Gu at al.~\cite{Gu} used 
the quantum entanglement between a local part and the rest of 
the system and the structure factor of charge-density wave (CDW) for heavy 
electrons.  Away from half-filling, they found that the domain of phase 
separation is not confined only to small $t_f$, but persists up to relative 
large values of $t_f$ (e.g., $t_f\sim 0.2$ for intermediate Coulomb
interactions $U\sim 5$) and with increasing $U$ is further stabilized. 
The same result has been obtained also by Wang et al. using the bosonization 
method~\cite{Wang}.

In the current paper we study the ground-state phase diagram of the 
asymmetric Hubbard model by an improved Lyzwa's scheme~\cite{Lyzwa1} discussed
above. The advantage of this method is that it can treat much larger clusters
than accessible by exact diagonalization technique and its applicability, 
unlike the DMRG~\cite{Gu} or bosonization~\cite{Wang} method, is 
not confined only to the one-dimensional case. Moreover, the method 
allows to calculate directly the probabilities of particular $f$ electron 
configurations and thus to specify the stability domains corresponding 
to distributions with the highest probabilities.

\section{Method}

Before discussing our approach, it is useful and instructive to summarize
main steps of the numerical algorithm leading to the exact solution of the 
spinless Falicov-Kimball model on finite clusters.  
The asymmetric Hubbard model (1) reduces to the spinless Falicov-Kimball 
model for $t_f=0$. Its Hamiltonian reads

\begin{equation}
H_{FKM}=-t_d\sum_{<ij>}d^+_id_j+U\sum_if^+_if_id^+_id_i.
\end{equation}

Since in this version, without $f$-electron hopping, the $f$-electron 
occupation number $f^+_if_i$ of each site $i$ commutes with
the Hamiltonian (2), the $f$-electron occupation number
is a good quantum number and can be replaced by classical variables 
$w_i=1$ or 0, according to whether or not the site $i$ is occupied
by the localized $f$ electron.
Then the Hamiltonian (2) can be written as
\begin{equation}
{\cal H_{FKM}}=\sum_{ij}h_{ij}d^+_id_j,
\end{equation}
where $h_{ij}(w)=-t_d$, if $i$ and $j$ are the nearest neighbor; 
$h_{ij}(w)=Uw_i$, if $i=j$ and zero otherwise.

Thus for a given $f$-electron configuration
$w=\{w_1,w_2 \dots w_L\}$ defined on the one, two, or three-di\-men\-sional
lattice, the Hamiltonian (3) is the second-quantized version of 
the single-particle Hamiltonian and can be directly diagonalized by 
the following canonical transformation
\begin{equation}
d^+_{\alpha}(w)=\sum_iV^{(w)}_{i\alpha}d^+_i\ ,
\end{equation}
where $V^{(w)}$ is the unitary matrix that diagonalizes $h(w)$. 
The ground state of ${\cal H_{FKM}}$ is then constructed
as follows
\begin{equation}
| \psi^{d}_{w}\rangle=\prod_{\alpha=1}^{N_d}d^+_{\alpha}(w)|0\rangle\ 
\end{equation}
and the corresponding ground-state energy is simply given by 
\begin{equation}
E_0(w)=\sum_{\alpha=1}^{N_d}\varepsilon_{\alpha}(w)\ , 
\end{equation}
where $\varepsilon_{\alpha}$ $(\varepsilon_1<\varepsilon_2<\dots <\varepsilon_L)$ 
are eigenvalues of the single particle matrix $h(w)$.

Generalizing this procedure our approach to the full Hamiltonian of the 
asymmetric Hubbard model (1) can be formulated in the following two points:
First, we construct the reduced basis $| \psi_k \rangle$ of $H$ by making
the Ansatz
\begin{equation}
| \psi_k \rangle=| \psi^f_k \rangle | \psi^d_k \rangle\ ,
\end{equation}
where $| \psi^f_k \rangle$ is the complete set of eigenstates of the
$f$-electron subsystem $(k=1,2,\dots ,\frac{L!}{(L-N_f)!N_f!})$ and 
$| \psi^d_k \rangle$ is the ground state  corresponding to 
$| \psi^f_k \rangle$ 
(note that $f^+_if_i| \psi^f_k \rangle$=$w^{(k)}_i| \psi^f_k \rangle$).
Second, the reduced basis $| \psi_k \rangle$ is used to calculate the matrix
elements $H_{nm}=\langle \psi_n |H| \psi_m\rangle$ of the full Hamiltonian
(1). This matrix is then diagonalized and its lowest energy eigenvalue $E_1$
yields the upper bound for the ground state energy of $H$. The corresponding
eigenvector
\begin{equation}
| \psi_G \rangle=\sum_nU_{n,1}|\psi_n\rangle\ ,
\end{equation}
(where $U_{nm}$ is the unitary matrix that diagonalizes $H_{nm}$) is the
approximate ground state and can be used directly to calculate the
expectation value of any operator 
$\langle \widehat{A} \rangle=\langle \psi_G|\widehat{A}|\psi_G \rangle$. 
Using expressions (4-8) one can show easily that any ground-state
expectation value can be written directly in terms of the unitary matrices
$V^{(w)}$ that diagonalize the single-particle matrices $h(w)$ corresponding
to all possible distributions of $f$-electrons.

To establish connection between our approach and Lyzwa's one let us
calculate explicitly the matrix elements $H_{nm}$ of the total Hamiltonian
of the asymmetric Hubbard model (1).
Making use of the fact that the new creation ($d^+_{\alpha}$) and annihilation
($d_{\alpha}$) operators obey the following commutation relation
\begin{equation}
d_{\alpha}(n)d^+_{\beta}(m)+d^+_{\beta}(m)d_{\alpha}(n)=k_{\alpha
\beta}(n,m)\ ,
\end{equation}
where the matrix elements $k_{\alpha \beta}$ are given by
\begin{equation}
k_{\alpha \beta}(n,m)=\sum_i {V_{i\alpha}^{(n)}}V_{i\beta}^{(m)}\ ,
\end{equation}
the straightforward calculations lead to the following expressions for the
diagonal and off-diagonal matrix elements of $H$
\begin{equation}
H_{nn}=E_0(n)
\end{equation}
and
\begin{equation}
H_{nm}=-t_f\langle \psi_n^f|\sum_{\langle ij\rangle}f^+_if_j |\psi_m^f \rangle
=-(-1)^{s}t_f\det(k(n,m))\ ,
\end{equation}
where $s$ is the number of permutations that should be done to transform
$\psi^f_m$ on $\psi^f_n$ by $f^+_if_j$.
As discussed bellow the Lyzwa's approach can be recovered directly from
Eq.~12 by putting $\det(k) =1$, however generally $\det(k)\neq1$.

To test our method we have first calculated the ground-state energy of the
asymmetric Hubbard model for various model parameters ($t_f$, $U$) on small
one-dimensional clusters where exact results are also accessible. 
In Table~1 and Table~2 we present the exact and approximate ground-state
energies obtained for the finite clusters of $L=6$ and 10 sites, two
representative values of $t_f$ ($t_f$=0.1, 1) and three representative
values of $U$ ($U$=0.1, 1, 10). For a comparison we have included into
Table~1 also results obtained by Lyzwa et al.~\cite{Lyzwa1} and we have
verified numerically that these results can be recovered exactly by 
our method simply putting $\det(k)=1$ in the Eq.~12. 
Comparing these results one can see that both approaches work very well in
the weak-coupling limit ($U\leq 1$), while in the opposite limit ($U\gg 1$)
our method yields a considerable better estimation of the ground-states
energy than the Lyzwa's one. Fig.~1 demonstrates that this trend holds also
for smaller values of $t_f$. Moreover, as one can expect intuitively, the
accordance between our results and the exact ones considerably improves with
decreasing $t_f$ and for a sufficiently strong asymmetry
($t_f \sim 0.2$) a nice accordance of results is observed over the whole
interval of Coulomb interactions (see also Table~2).

To verify the ability of our method to describe the main characteristics of
the exact ground state we have also calculated the $f$-electron pair
correlation function $L(x)$ defined by~\cite{Lyzwa1} 
\begin{equation}
L(x)=\frac{1}{L}\sum_j \langle f^+_jf_jf^+_{j+x}f_{j+x}\rangle.
\end{equation}
The results of numerical computations are summarized in Fig.~2. It is seen
that a nice accordance with exact behaviors is obtained in the weak ($U=1$) 
as well as strong ($U=10$) coupling limit for both small ($t_f\leq 0.2$) 
and intermediate ($t_f\sim 0.4$) values of $f$-electron hopping integrals.

\section{Phase diagrams}
\subsection{One-dimensional case}
One of the greatest advantages of the method discussed above is that it allows 
to calculate directly the probability of any $f$-electron configuration
in the approximate ground state. We used this fact to construct the phase
diagrams of the asymmetric Hubbard model in the $t_f-U$ plane for various
sizes of clusters and electron fillings. The phases presented in the phase
diagrams are those corresponding to $f$-electron distributions with the highest 
probabilities for given values of $t_f$ and $U$.

The typical examples of ground-state phase diagrams of the asymmetric
Hubbard model are displayed in Fig.~3 for $L=24$ and four representative 
values of $f$-electron filling $n_f$ (note that $n_d=n_f=n/2$).
A general feature which can be noticed in these pictures is that the basic
structure of the phase diagrams is formed by only two main types of the
$f$-electron configurations, and namely, the most homogeneous distributions
$MHD$ (the $f$ electrons are distributed homogeneously over the whole lattice) 
and the phase separated configurations (the $f$ electrons occupy only one part
of the lattice while the remaining one is empty). As discussed below, 
between the phase separated configurations the special role play the phase 
segregated configurations (all $f$ electrons clump together) and therefore
they are considered here as the independent group. In the language of an
effective interaction between the $f$ electrons, the most homogeneous
configurations correspond to an effective repulsion and the phase 
separated/segregated configurations to an effective attraction between 
the $f$ electrons. From this point of view, it is very interesting that 
in the pure electronic 
system (with only the on-site Coulomb repulsion between the light and heavy
electrons) an effective attraction between the $f$ electrons is produced,
even for $t_f$ away from the Falicov-Kimball limit $t_f=0$. Indeed, our
results show that the phase boundary $t^c_f(U)$ between the most homogeneous 
and phase-separated/segregated regions increases rapidly with increasing $U$
and reaches the intermediate values $t^c_f(U)\sim 0.25$ already for
intermediate Coulomb interactions. In the weak coupling and low density 
limit, the phase boundary scales like $t^c_f(U)\sim U^2$, while at higher 
electron fillings ($n_f=1/4$ and $n_f=1/3$) the phase-separated/segregated 
distributions are stabilized above some critical value of Coulomb
interaction $U$ ($U\sim 0.5$ for $n_f=1/4$ and $U\sim 2.2$ for $n_f=1/3$).
In these limiting cases our results reproduce the analytical and numerical
results obtained recently by bosonization~\cite{Wang} and exact-diagonalization/DMRG
method~\cite{Gu}. Comparing our results with the exact-diagonalization/DMRG 
results~\cite{Gu} ($n_f=1/4$) one can see that these results agree very well
in spite of the fact that  fully different approaches have been used to 
identify the phase separated region.

The advantage of our method is that it also allows us to identify the internal
structure of the phase diagrams and thereby to study how this structure 
changes by varying the model parameters. Fig.~3 shows that the phase
diagrams of the asymmetric Hubbard model (strictly said the phase-separated
domains) have a rich internal structure that exhibits some general trends.
First, the phase separated region starts with the phase segregated
distribution. Small exceptions are found only for $n_f=1/4$ and $n_f=1/3$,
where also some other phases are observed for $t_f\rightarrow 0$, but their
stability regions are very limited. 
Second, the segregated cluster of length $N_f$ splits into
two or more smaller clusters with increasing $t_f$. Third, the segregated
configuration is stabilized with increasing electron filling, while the
separated phases disappear from the phase diagrams.
 
Although the cluster used in our numerical calculations is relatively large
($L=24$) to exclude completely the finite size effects the same calculations
have been performed on several different clusters for each selected value of
electron filling. We have found that the fundamental characteristics of the
phase diagrams discussed above, and namely, the phase boundary between the
phase-separated and most homogeneous phase, the phase boundary of segregated
phase, the critical value of Coulomb interactions at which the phase
separation starts for large $f$-electron fillings are almost independent of
$L$. This is clearly demonstrated in Fig.~4 where the boundary of phase
separation is plotted for two different values of $n_f$
($n_f=1/4$ and $n_f=1/3$). This analysis indicates that our one-dimensional 
results can be extrapolated satisfactorily to the thermodynamic 
limit ($L\rightarrow \infty$).

\subsection{Two-dimensional case}
Since the phenomenon of phase separation is one of the most interesting
problems in the condensed matter physics we extend our calculations also on the 
two-dimensional case. As discussed above such an extension is possible 
due to the fact that instead the full Hilbert space we work only 
with the reduced basis and corresponding reduced matrices of rank
$\frac{L!}{(L-N_f)!N_f!}$. This allows us to study the two-dimensional clusters
up to $6\times 6$ sites that are far away beyond the reach of present day
computers within the exact diagonalization calculations. The results of our
numerical calculations obtained for two different values of electron fillings 
are summarized in Fig.~5 in the form of $t_f-U$ phase diagrams together
with the complete list of $f$-electron configurations with the highest
probabilities. It is seen that all main features of the one-dimensional 
phase diagrams hold in two dimensions, too. For small values of $f$-electron 
hopping the system is phase segregated/separated, while increasing
$t_f$ stabilizes the homogeneous distribution of $f$ electrons. 
In accordance with the one-dimensional case we have found that 
the phase boundary $t_f^c(U)$ between the homogeneous and phase separated
region scales like $U^2$ for weak interactions, and that the region of 
phase segregation/separation increases rapidly with increasing $n_f$.  

In summary, we have presented an improved numerical scheme for calculating 
ground-state properties of the asymmetric Hubbard model.  The advantage of 
this method is that it can treat much larger clusters than accessible by 
exact diagonalization technique and its applicability, unlike 
the DMRG or bosonization method, is not confined only
to the one-dimensional case. Moreover, the method allows to calculate 
directly the probabilities of particular $f$ electron configurations
and thus to specify the stability domains corresponding to distributions
with the highest probabilities. We have used this fact to construct
the ground-state phase diagrams of the asymmetric Hubbard model in one 
and two dimensions for wide range of model parameters. We have found that 
at low electron fillings ($n\leq 1/3$) the ground states are always phase
segregated for a strong asymmetry between the hopping integral 
of $d$ and $f$ electrons ($t_d\gg t_f$). With decreasing asymmetry 
the system undergoes a transition to the phase separated state and
then to the homogeneous state. For electron fillings $n>1/3$  and 
weak Coulomb interactions the ground state (in one dimension)
is homogeneous for all values of asymmetry, while for intermediate and
strong interactions the system exhibits the same sequence of phase
transitions as for $n$ small.

\vspace{0.5cm}
This work was supported by the Slovak Grant Agency for Science
under grant No. 2/7057/27 and the Slovak APVV Grant Agency under Grant
LPP-0047-06. I would also like to acknowledge H. \v Cen\v carikov\'a for a 
technical help during the preparation of manuscript.

\newpage

\begin{table}[h]
\caption{The ground-state energy of the one-dimensional asymmetric Hubbard 
model calculated for three different values of Coulomb interaction $U$ on 
finite clusters of $L=6$ and $L=10$ sites at $t_f=1$ and $n_f=n_d=1/2$.
Different columns correspond to exact results (Exact), Lyzwa's approach
(Approx.~I) and our approach (Approx.~II).}
\begin{center}
\begin{tabular}{c|c|c|c|c|c|c}
\hline\hline
     & 
      \multicolumn{3}{c|}{L=6}         &   \multicolumn{3}{c}{L=10} \\ \cline{2-7}
U    & Exact & Approx.~I & Approx.~II   & Exact & Approx.~I & Approx.~II\\
\hline
0.1  & -1.30850 & -1.30868  & -1.30830   & -1.26960 & -1.26978  & -1.26934 \\
1.0  & -1.10019 & -1.11770  & -1.08064   & -1.06144 & -1.08013  & -1.03697 \\
10   & -0.27739 & -0.73823  & -0.20421   & -0.27037 & -0.71753  & -0.19908 \\
\hline\hline
\end{tabular}
\end{center}
\end{table}

\begin{table}[h]
\caption{The ground-state energy of the one-dimensional asymmetric Hubbard 
model calculated for three different values of Coulomb interaction $U$ on 
finite clusters of $L=6,10$ and $L=14$ sites at $t_f=0.1$ and $n_f=n_d=1/2$.
Different columns correspond to exact results (Exact) and our approach 
(Approx.~II).}
\begin{center}
\begin{tabular}{c|c|c|c|c|c|c}
\hline\hline
     & 
  \multicolumn{2}{c|}{L=6} & \multicolumn{2}{c|}{L=10} &  \multicolumn{2}{c}{L=14} 
\\ \cline{2-7}
U     & Exact & Approx.~II   & Exact & Approx.~II & Exact & Approx.~II \\
\hline
0.1  & -0.70864 & -0.70864  &  -0.68725  & -0.68725 & -0.68152  & -0.68151 \\
1.0  & -0.51544 & -0.51507  &  -0.49546  & -0.49495 & -0.49047  & -0.48986 \\
10   & -0.10056 & -0.09961  &  -0.10015  & -0.09925 & -0.10018  & -0.09923 \\
\hline\hline
\end{tabular}
\end{center}
\end{table}

\newpage
Figure Captions

\vspace{0.5cm}
Fig.~1. The ground-state energy of the asymmetric Hubbard model as a function 
of $U$ calculated for four different values of $t_f$ and $L=10$. Different 
lines correspond to exact results (solid line), our approach (dashed line) 
and Lyzwa's approach (dashed-dotted line). 
The half-filled band case ($n_f=n_d=1/2$).

\vspace{0.5cm}
Fig.~2. The $f$-electron pair correlation function $L(x)$ of the asymmetric 
Hubbard model calculated for two different values of $U$ and $t_f$ at $L=10$. 
Different lines correspond to exact results (solid line), our approach 
(dashed line) and Lyzwa's approach (dashed-dotted line).
The half-filled band case ($n_f=n_d=1/2$).

\vspace{0.5cm}
Fig.~3. The ground-state phase diagram of the one-dimensional asymmetric 
Hubbard model calculated for several $f$-electron densities on finite 
cluster of $L=24$ sites. Depicted domains represent the stability regions 
of the $f$-electron configurations with the highest probabilities in the 
approximate ground state. Two small domains $w_a$ and $w_b$ (for $n_f=1/3$) 
consists of several subdomains: $w^1_a=[110000]_4$, $w^2_a=[110011000000]_2$;
$w^1_b=[1100]_400000000$, $w^2_b=111001100111000000000000$, 
$w^3_b=[111100]_2000000000000$, where the
lower index denotes the number of repetitions of the block $[\dots]$.

\vspace{0.5cm}
Fig.~4. The phase boundary between the homogeneous (MHD) and phase separated
(PS) domain as a function of $U$. Comparison of
our numerical results obtained on different finite clusters for $n_f=1/4$
and $n_f=1/3$.

\vspace{0.5cm}
Fig.~5. The ground-state phase diagram of the two-dimensional asymmetric 
Hubbard model calculated
for two $f$-electron densities on finite cluster of $L=36$ sites with the
complete list of $f$-electron configurations  with the highest probabilities.

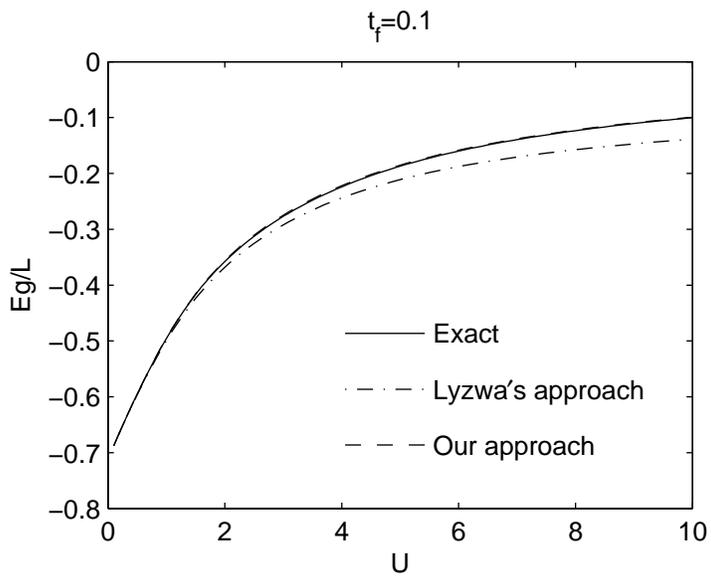
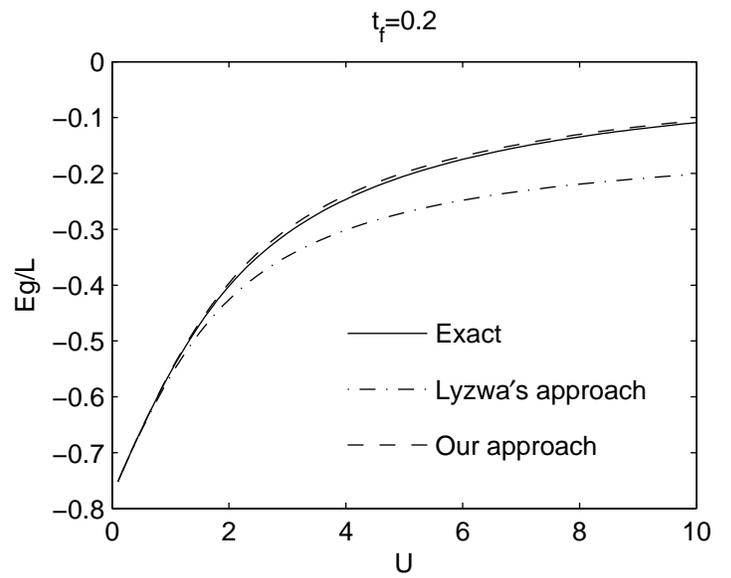
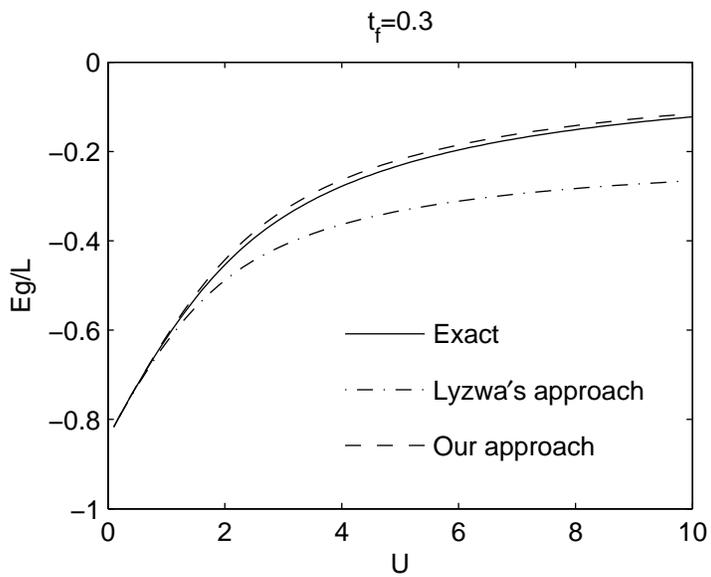
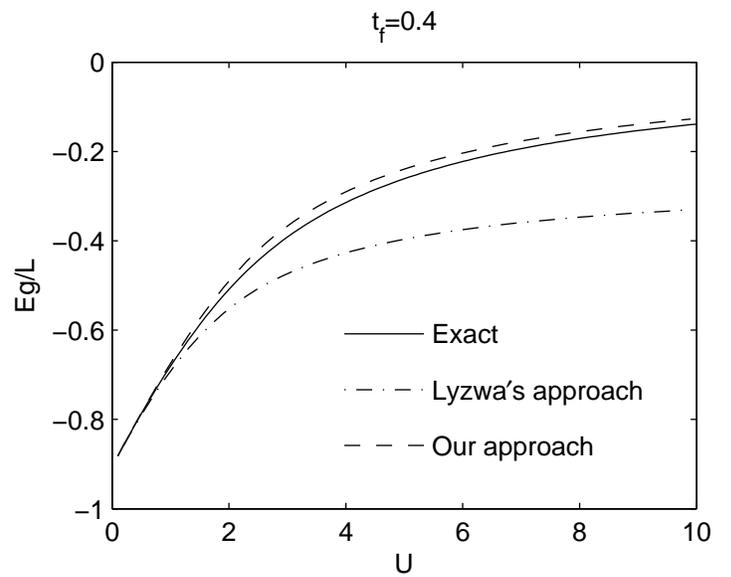
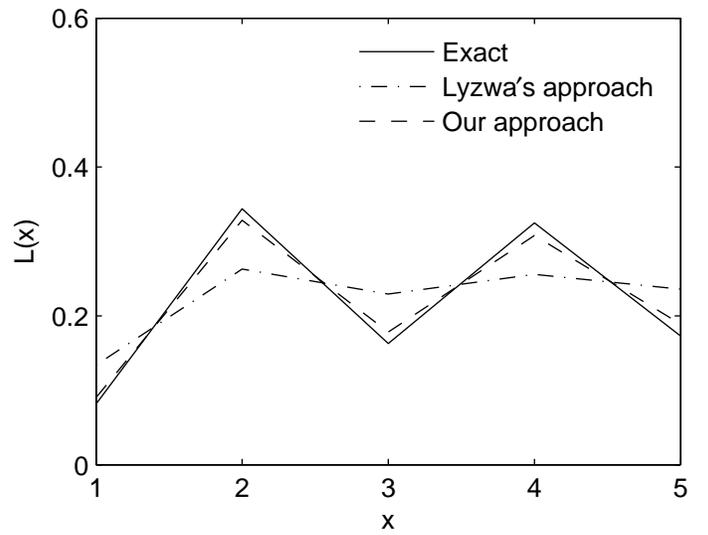
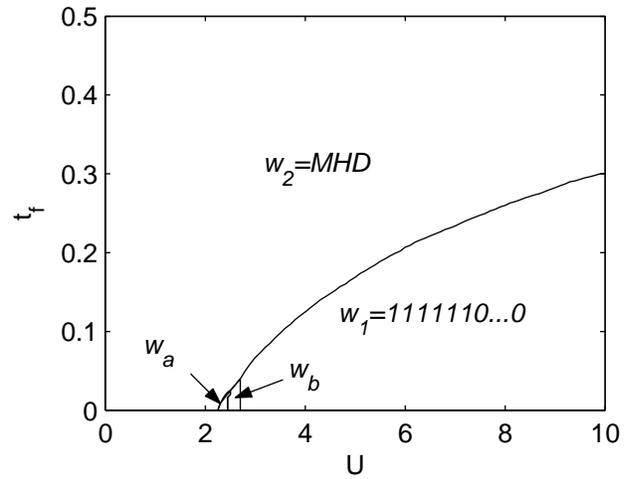
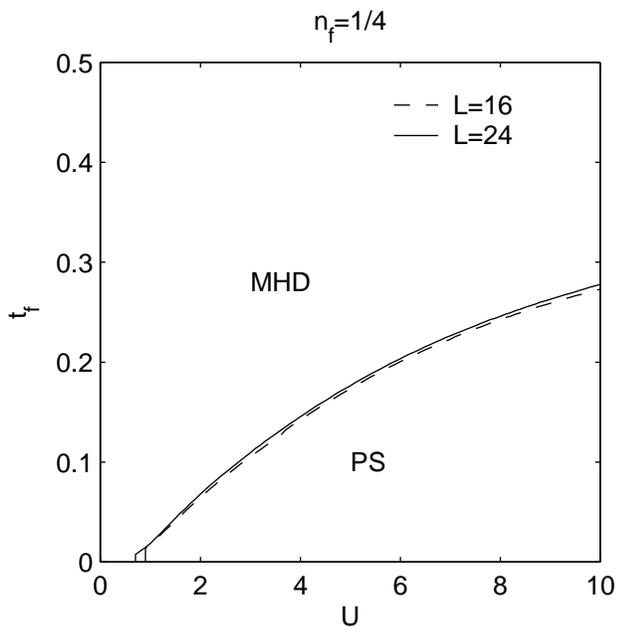
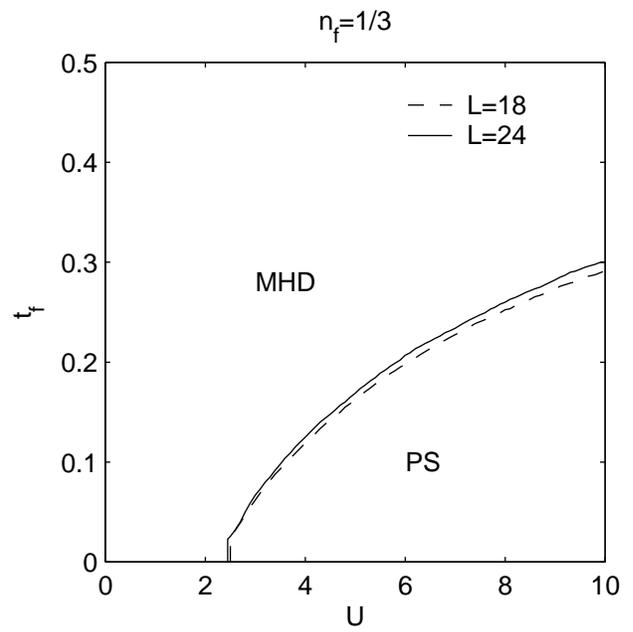
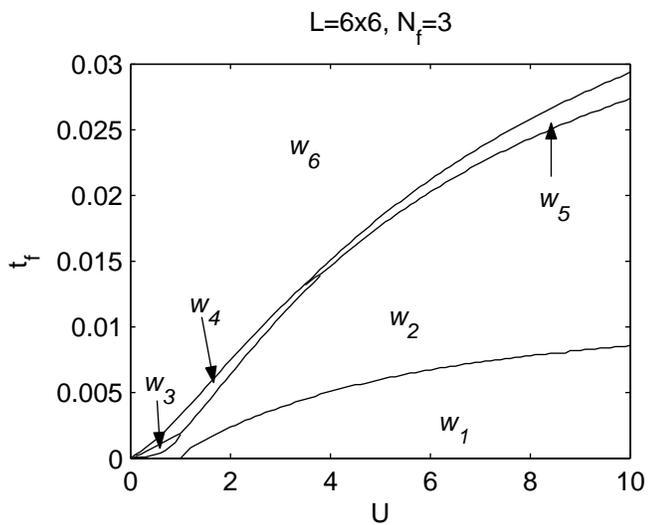
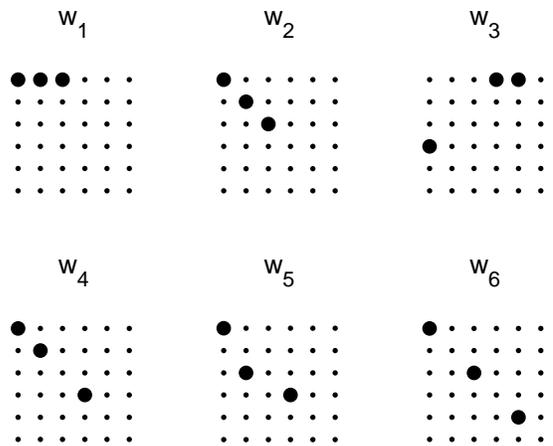
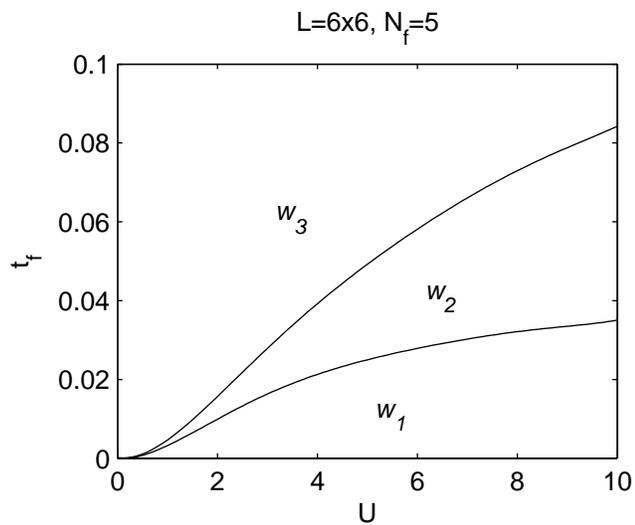
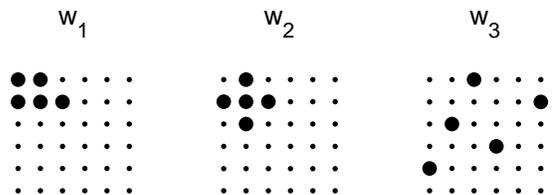
\end{document}